\documentclass{mem}
\usepackage{natbib}\usepackage{txfonts}\usepackage{balance}
\usepackage{graphicx}
\usepackage[a4paper]{hyperref}
\idline{80}{882}
\begin{document}

\newcommand{\dmu}{\Delta\mu/\mu}
\newcommand{\lya}{Ly$\alpha$ }
\newcommand{\kms}{${\rm km\,s}^{-1}$}

\def \epjst{Eur.~Phys.~J-Spec.~Top.}
\def \jms{J.~Molecular~Spectrosc.}
\def \al{Astron.~Lett.}
\def \spjetpl{Sov.~Phys.~JETP~Lett.}

\title{
New limit on a varying proton-to-electron mass ratio from high-resolution optical quasar spectra
}

	\subtitle{}

\author{
A.~L.~Malec\inst{1} 
\and R.~Buning\inst{2}
\and M.~T.~Murphy\inst{1}
\and N.~Milutinovic\inst{3}
\and S.~L.~Ellison\inst{3}
\and J.~X.~Prochaska\inst{4}
\and L.~Kaper\inst{2,5}
\and J.~Tumlinson\inst{6}
\and R.~F.~Carswell\inst{7}
\and W.~Ubachs\inst{2}
       }

	\offprints{A.~L.~Malec}

\institute{
Centre for Astrophysics and Supercomputing, Swinburne University of Technology, Melbourne, Victoria 3122, Australia,
\email{amalec@swin.edu.au}
\and
Laser Centre, VU University, De Boelelaan 1081, 1081 HV Amsterdam, The Netherlands
\and
Department of Physics and Astronomy, University of Victoria, Victoria, BC, V8P 1A1, Canada
\and
University of California Observatories -- Lick Observatory, University of California, Santa Cruz, CA 95064
\and
Astronomical Institute Anton Pannekoek, Universiteit van Amsterdam, 1098 SJ Amsterdam, The Netherlands
\and
Yale Center for Astronomy and Astrophysics, Department of Physics, New Haven, CT 06520, USA
\and
Institute of Astronomy, University of Cambridge, Madingley Road, Cambridge, CB3 0HA, UK
}

\authorrunning{Malec }
\titlerunning{Limit on the variation of the proton-to-electron mass ratio}

\abstract{Molecular transitions recently discovered at redshift
  $z_{\rm abs}=2.059$ toward the bright background quasar J2123$-$0050
  are analysed to limit cosmological variation in the
  proton-to-electron mass ratio, $\mu\equiv m_{\rm p}/m_{\rm e}$.
  Observed with the Keck telescope, the optical spectrum has the
  highest resolving power and largest number (86) of H$_2$ transitions
  in such analyses so far.  Also, (7) HD transitions are used for the
  first time to constrain $\mu$-variation. These factors, and an
  analysis employing the fewest possible free parameters, strongly
  constrain $\mu$'s relative deviation from the current laboratory
  value: $\dmu =(+5.6\pm5.5_{\rm stat}\pm2.7_{\rm sys})\times10^{-6}$.
  This is the first Keck result to complement recent constraints from
  three systems at $z_{\rm abs}>2.5$ observed with the Very Large
  Telescope.
}
\maketitle{}

\section{Introduction}\label{sec:intro}

The assumed invariance of fundamental constants within the Standard Model of particle physics must be tested experimentally. Temporal or spatial deviations from the currently measured laboratory values of important constants, such as the fine structure constant, $\alpha$, and the proton-to-electron mass ratio, $\mu\equiv m_{\rm p}/m_{\rm e}$, could point to more fundamental theories of physical interactions. 

Limits on the variation of $\mu$ have been obtained in Earth-bound laboratory experiments \citep[e.g.][]{BlattS_08a} but can also be derived over larger cosmological scales from molecular hydrogen absorption in the spectra of quasars \citep{ThompsonR_75a,VarshalovichD_93b}. Variation of $\mu$ manifests in a mass-dependent velocity shift, $\Delta v_i$, of a ro-vibronic transition $i$. The magnitude and direction of $\Delta v_i$ is characterised by a sensitivity coefficient $K_i$. Measurement of a shift in the observed redshift of the transition, $\Delta z_i$, relative to the redshift of the absorption cloud, $z_{\rm abs}$, in which the transition is observed, implies a variation of $\mu$,
\begin{equation}
	\frac{\Delta v_i}{c} \approx \frac{\Delta z_i}{1+z_{\rm abs}} = K_i \frac{\Delta\mu}{\mu}\,, \label{eq:shifts}
\end{equation}
where $\Delta\mu/\mu\equiv(\mu_z-\mu_{\rm lab})/\mu_{\rm lab}$ for $\mu_{\rm lab}$, the current laboratory value of $\mu$ and $\mu_z$ the value in the absorption cloud. Using two or more transitions allows for $\dmu$ and $z_{\rm abs}$ to be simultaneously measured.

Current studies derive limits on $\dmu$ using a small sample of H$_2$ absorption systems from the UVES instrument on the Very Large Telescope (VLT) in Chile \citep{IvanchikA_05a, ReinholdE_06a}. More recent analyses, also VLT-based, yield null constraints \citep{KingJ_08a, WendtM_08a, ThompsonR_09a}. 

In this paper we describe our analysis of a new H$_2$ \& HD absorber observed with the Keck telescope. Full details of the analysis will be presented in \citet{MalecA_10a}.

\section{Data}\label{sec:data}

\subsection{Keck spectrum of J2123$-$0050}\label{ssec:keck}

We study the newly discovered H$_2$ absorber at $z_{\rm abs}=2.059$ towards the $z_{\rm em}=2.261$ quasar SDSS J212329.46$-$005052.9 (hereafter J2123$-$0050; Milutinovic et al., in preparation). Given the scarcity of such systems \citep{NoterdaemeP_08a} and further selection criteria imposed by varying-$\mu$ analyses ($z_{\rm abs}>2$ to shift enough transitions above the atmospheric cutoff, high enough H$_2$ column densities and background quasar luminosities) this new absorber is exceptional. The $R\approx110000$ spectrum was obtained using the HIRES instrument on the Keck telescope in Hawaii and covers wavelengths 3071--5896\,\AA. We find 86 H$_2$ and, for the first time in such analyses, 7 HD transitions usable for constraining $\dmu$. Notably, this is also the first time Keck/HIRES-based data is used to constrain $\mu$-variation with similar precision to the previous VLT results.
\begin{figure}[t!]
\begin{center}
\includegraphics[width=0.45\textwidth]{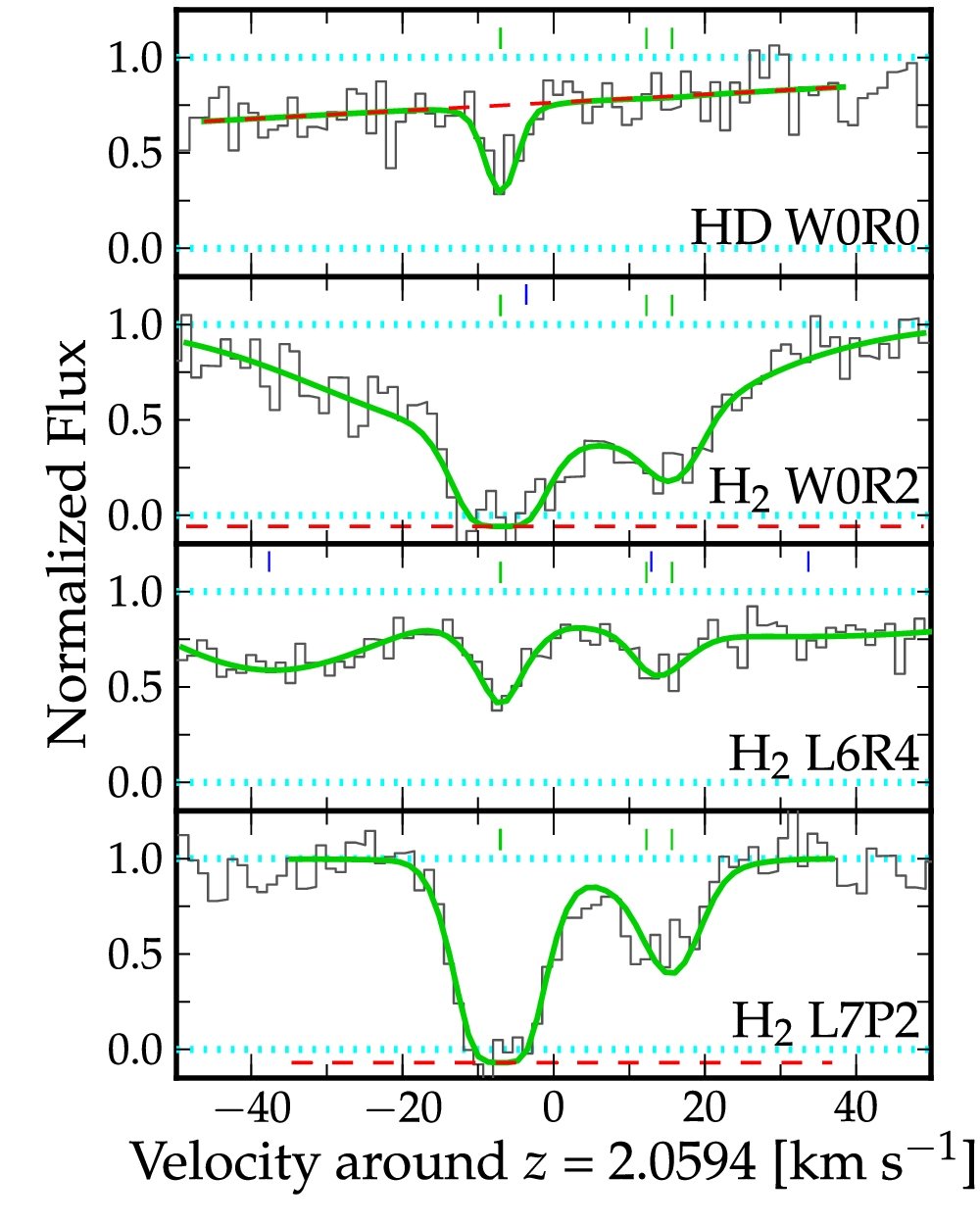}\vspace{-0.5em}
\caption{\footnotesize
Some of the 86 H$_2$ and 7 HD lines from the J2123$-$0050 Keck spectrum on a velocity scale centred at $z_{\rm abs}=2.0594$. The molecular lines fall in the wavelength range 3100--3421\,\AA\ with SNR 7--25 per 1.3\,\kms\ pixel. The spectrum (black histogram) is normalized by a nominal continuum (upper dotted line). Also shown are the fiducial 4 component fit (solid curve, VCs marked by tick marks), local continua and zero levels (upper and lower dashed lines) and \lya lines (denoted by tick marks offset above the molecular VCs). Note that the two left-most components are nearly coincident in velocity.
}
\label{fig:trans}
\end{center}
\end{figure}

Figure \ref{fig:trans} shows a small, but representative subset of H$_2$ and HD transitions in the data, with a range of overall line-strengths, SNRs and ground-state rotational levels, characterised by the quantum number $J$ (`$J$-levels'). Two distinct spectral features (SFs) are seen, separated by $\approx$20\,\kms. This is distinct from the velocity components (VCs) used to model the molecular absorption system. In Section \ref{ssec:fiducial} we find evidence for 4 VCs given the SNR of the data. The strong left-hand SF appears saturated for most low-$J$ transitions and the weaker right-hand SF appears unsaturated in almost all transitions. Only the left-hand SF is detected in the HD transitions.

\subsection{Molecular data}\label{ssec:moldat}


Quasar-based varying-$\mu$ analyses rely on sub-pixel measurements of deviations of astronomical transition wavelengths from laboratory wavelengths, and on the sensitivity coefficients, $K_i$, that characterise the deviations. For H$_2$, we use the most accurate laboratory wavelengths available from \citet{AbgrallH_93c}, \citet{UbachsW_07a} and \citet{SalumbidesE_08a}. For HD, we use the wavelengths listed in \citet{IvanovT_08a}. The H$_2$ and HD $K_i$ coefficients used were calculated by \citet{UbachsW_07a} and \citet{IvanovT_08a}, respectively. 

\section{Analysis and Results}\label{sec:analysis}

\subsection{Fit and $\chi^2$ minimization analysis}\label{ssec:chimin}

Previous analyses \citep[e.g.][]{IvanchikA_05a, ThompsonR_09a} used a `line-by line' fitting approach. The redshifts of individual H$_2$ lines were derived from independent fits. Using a linear fit to the reduced redshifts versus the $K_i$ coefficients, the value of $\dmu$ can be derived as defined in equation (\ref{eq:shifts}). This approach requires that only a single VC is fit to the molecular lines. Lines blended with neighbouring features were excluded from these analyses.

All of the molecular lines of interest in our data fall in the \lya forest. Many are blended (to varying degrees) with \lya absorption lines. Some are blended with metal lines in the same absorption system (we model this absorption accurately using transitions that fall outside of the \lya forest). The molecular absorption itself shows two SFs which, while distinct, are `blended'. To utilise the many H$_2$ lines present in the data, a number of which would otherwise be rejected, H$_2$ and HD absorption profiles, together with the broader Lyman-$\alpha$ profiles, are fit simultaneously. With a lesser influence on the derived value of $\dmu$, there also exist local uncertainties in the zero and continuum flux levels, set as free parameters in the fit. This `simultaneous fitting' technique allows $\dmu$ to be calculated with the maximum precision available in the data, while taking into account the uncertainties brought about by blends, level adjustments and also varying SNRs \citep[][describes further advantages of this technique]{KingJ_08a}.

Each line, or VC, associated with an absorbing cloud, including \lya and metal lines, is modelled as a Voigt profile, parametrized by the Doppler width, $b$, column density, $N$, and redshift, $z_{\rm abs}$. Different molecular transitions share these parameters in physically meaningful ways. For H$_2$ and HD we assume a common velocity structure: a given VC shares the same $z_{\rm abs}$ in all transitions. Each VC has a common $N$ in the same $J$-level and a common $b$ for all $J$-levels. Because HD lines are few and relatively weak the HD $N$ values are constrained to follow the same pattern across the profile as the H$_2$ $J$=0 values. In relating the absorption cloud parameters the number of free parameters in the absorption model is minimized. We relax these assumptions as part of the internal consistency checks described in Section \ref{ssec:sys} and find no evidence of strong deviations from our final result. The parameter of key interest, $\dmu$, is common to all molecular VCs. 

The $\chi^2$ between the data and the absorption model is minimised using the program {\sc vpfit}, designed for modelling complex, interrelated absorption lines with the ability to link physically related parameters as defined in our model. Statistical 1-$\sigma$ uncertainties on best fitting parameters, including $\dmu$, are derived by {\sc vpfit} from the final parameter covariance matrix. The reliability of the calculated uncertainty in $\dmu$   is tested using fits to simulated spectra in Section \ref{ssec:sim}.

\subsection{Fiducial absorption model}\label{ssec:fiducial}
\begin{figure*}[t!]
\begin{center}
\includegraphics[width=0.75\textwidth]{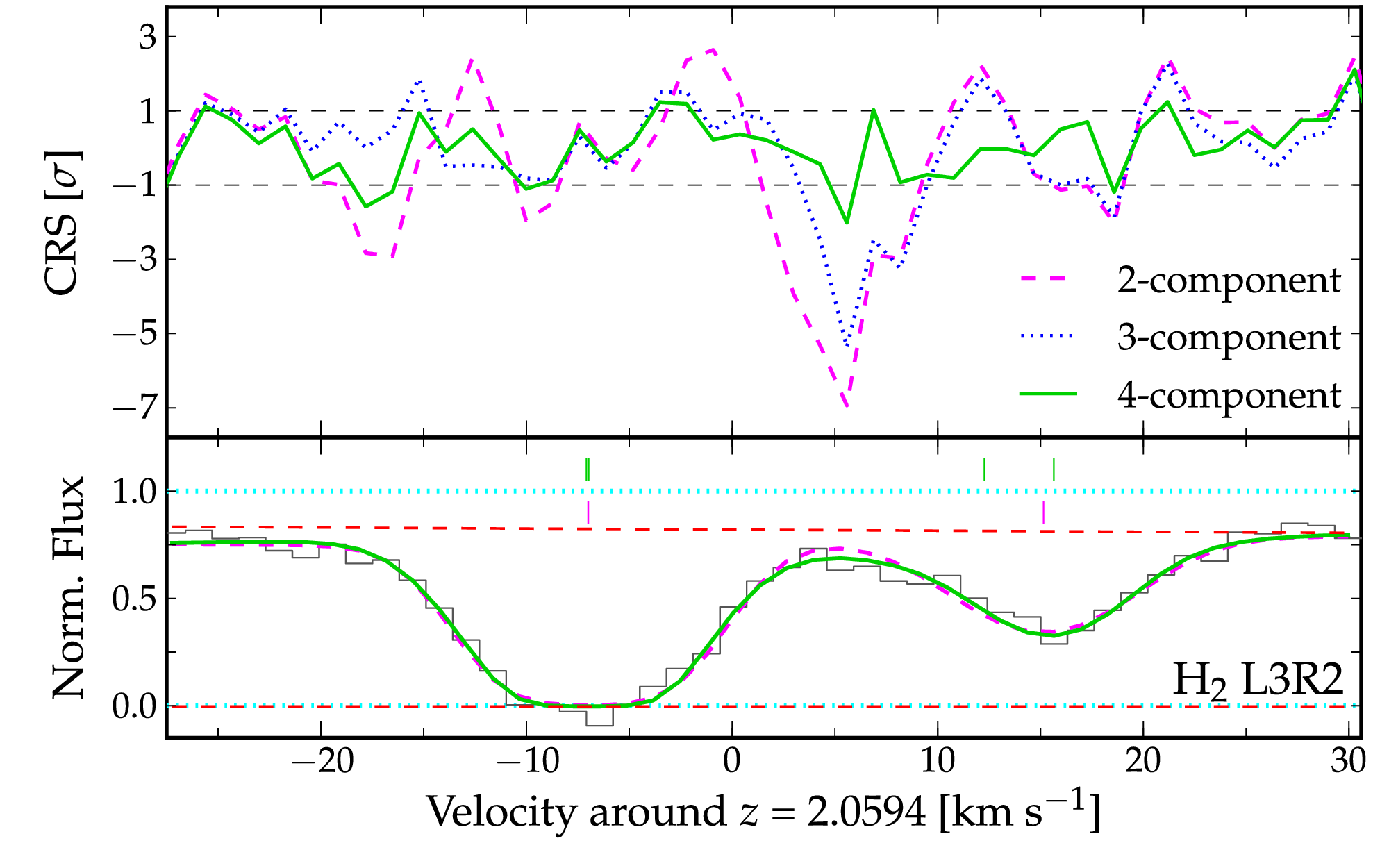}\vspace{-0.5em}
\caption{\footnotesize
Composite residual spectra (CRS), formed from 23 relatively unblended H$_2$ transitions. \emph{Bottom panel}: An example H$_2$ transition fitted with the 2 and fiducial 4 component models (dashed and solid curves, respectively). Only a small difference is noticeable by eye. \emph{Top panel}: CRS for the 2, 3 and fiducial 4 component models (dashed, dotted and solid lines respectively). The significant, many-pixel excursions outside the $\pm$1-$\sigma$ range for the 2 and 3 component models indicate unmodelled velocity structure, not readily apparent when inspecting the transitions ``by eye''.
}
\label{fig:crs}
\end{center}
\end{figure*}
The initial model for the molecular absorption used two VCs, one for each SF. The \lya forest lines, blending metal lines, continuum and zero levels were added and refined assuming this preliminary molecular model. The velocity structure was progressively built by adding molecular VCs. After each fit was optimised, the model with the lowest $\chi^2$ per degree of freedom, $\chi^2_\nu$, was selected as the fiducial one. The fiducial 4-component model was found to have a lower $\chi^2_\nu$ than the 3-component models, which in turn had lower $\chi^2_\nu$ values than the 2-component model. Fits using 5 and 6 VCs were attempted but the additional components were deemed statistically unnecessary by {\sc vpfit}.

We introduce here the `composite residual spectrum' (CRS) as another, complimentary tool for discriminating between different absorption models. It is constructed by averaging the 1-$\sigma$ normalized residual spectra (i.e.~$[{\rm data} - {\rm fit}]/{\rm error}$) of regions containing absorption lines of interest on a common velocity scale. A CRS is a better visual diagnostic of poorly fit velocity structure, especially under-fitting, than ``by eye'' inspection of a large number of H$_2$ transitions. It can be used to aid the construction of new fits. Figure \ref{fig:crs} shows a comparison of CRS plots for 2, 3 and 4 component fits. It is clear that the 2 and 3 component models fail to reproduce the statistical structure of the profile, leaving many-pixel excursions outside the expected residual range. Additional VCs significantly diminish these deviations, with no obvious evidence for unmodelled, statistically significant structure in the 4 VC fit.

The value of $\dmu$ is simply one of the free parameters in the fiducial 4 component absorption model for which $\chi^2$ has been minimized:
\begin{equation}\label{eq:stat_result} 
	\dmu = (+5.6\pm5.5_{\rm stat})\times10^{-6}. 
\end{equation}

\subsection{Simulations of $\chi^2$ minimization}\label{ssec:sim}

To verify the uncertainties on $\dmu$ returned by {\sc vpfit} we performed a Monte Carlo simulation of the fiducial 4 component absorption model used to derive our constraint. Every absorption line and fitting region in our model is used as the basis of a simulated spectrum. In each of 420 realisations the original flux error array was used to generate Gaussian noise in the simulated spectrum. Note that the error array was scaled by a factor of 0.8 to increase the statistical stability of the fits. The molecular lines in the simulated spectra use a $\dmu$ value of $5\times10^{-6}$ to verify that {\sc vpfit} is capable of detecting a non-zero deviation in $\mu$ from a zero starting point. Our fit is one of the largest and most complex constructed in a varying-$\mu$ study, with a large number of fitted pixels, free parameters, various links between parameters, zero and continuum level adjustments contributing to the final measurement of $\dmu$. This is the first time {\sc vpfit} has been tested in this computationally-expensive regime.

\begin{figure}[t!]
\begin{center}
\includegraphics[width=0.45\textwidth]{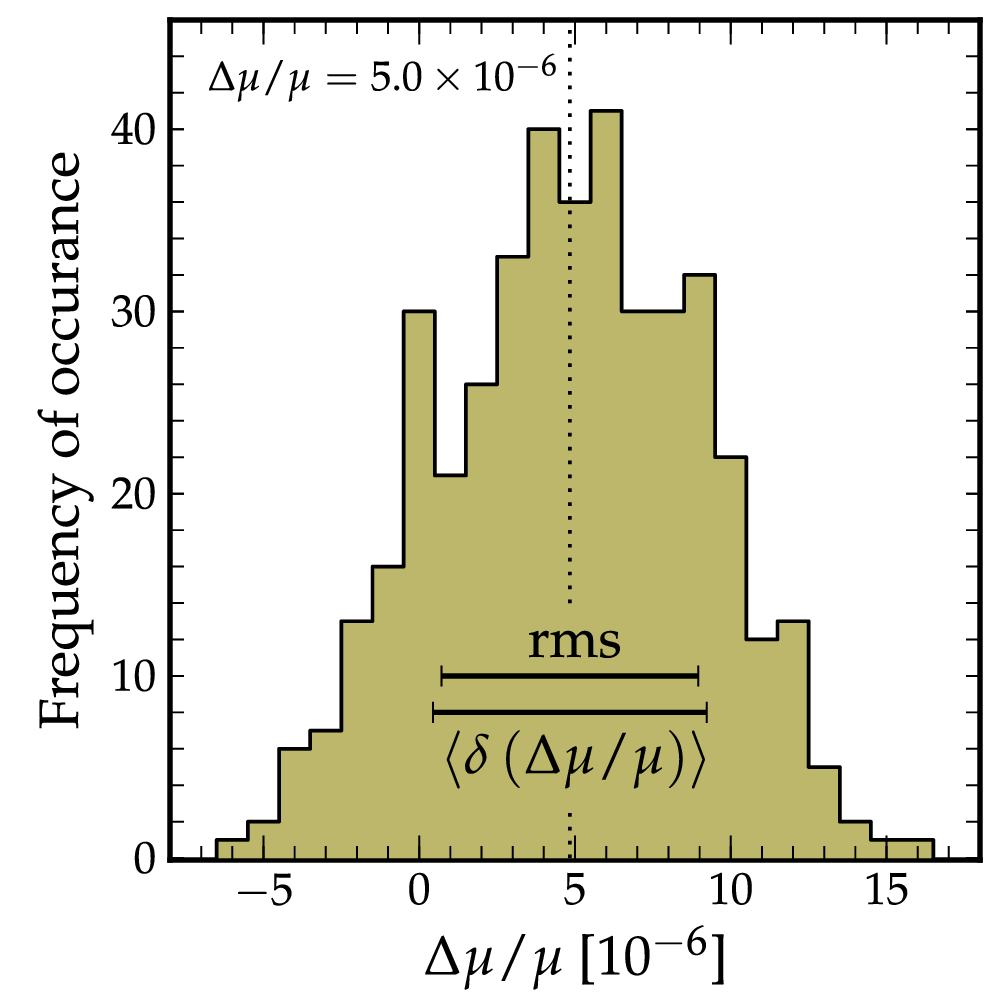}\vspace{-0.5em}
\caption{\footnotesize 
420 Monte Carlo simulations of the fiducial 4 component absorption model used in our analysis -- verification that {\sc vpfit} returns reliable, if not conservative, uncertainties. A variation in $\mu$ of $\dmu = 5\times10^{-6}$ was input into the simulations. The mean 1-$\sigma$ error and the standard deviation (labelled as rms) are also shown and are found to be consistent. The mean $\dmu$ value retrieved (vertical dotted line) corresponds well to the value input in the simulation.
}
\label{fig:sims}
\end{center}
\end{figure}

The results are plotted in Figure \ref{fig:sims}. The mean 1-$\sigma$ uncertainty derived on individual $\dmu$ measurements is $4.4\times10^{-6}$ (the expected value given a scaling of 0.8 of the flux errors). This corresponds well to the standard deviation of $4.1\times10^{-6}$ for the 420 $\dmu$ measurements and is consistent with the findings of \citet{KingJ_09a} in that the uncertainties returned by {\sc vpfit} may be slightly conservative. The mean $\dmu$ value retrieved is $4.8\times10^{-6}$.

\subsection{Potential systematic errors}\label{ssec:sys}

A wide range of internal consistency checks were run: only relatively `unblended' transitions were fit; the contribution of certain subsets of lines (higher $J$-levels, HD lines, Werner transitions) to $\dmu$ was turned off; oscillator strengths of all transitions were set to be free parameters; separate $\dmu$ parameters were fit for different $J$-levels; $b$ parameters were allowed to very between different $J$-levels. Some of these checks relaxed the assumptions made in Section \ref{ssec:chimin}. We found no evidence to indicate any obvious problems with our data or analysis. Details of these and other tests conducted will be described in \citet{MalecA_10a}.

Many H$_2$ and HD transitions are used, over a wide wavelength range, so the effect of calibration errors in the wavelength scale is substantially reduced. However, systematic errors due to long- or short-range calibration distortions are possible and were investigated. 

ThAr lamp exposures were used to calibrate our science exposures. We use the technique employed in \citet{MurphyM_07b} designed to estimate potential long-range distortions of the wavelength scale. The possible wavelength distortion in the region of molecular absorption was found to be $<30$\,m\,s$^{-1}$, corresponding to a systematic error in $\Delta\mu/\mu$ of $\pm2.0\times10^{-6}$ at most. 

A distinctly different type of distortion was recently identified by \citet{GriestK_10a} in the HIRES instrument. Comparisons of ThAr and iodine cell exposures revealed that transitions at echelle order edges appear at negative velocities with respect to transitions at the order centres when calibrated with a ThAr exposure. A calculation of the possible effect on $\dmu$ was obtained by introducing a `counter-distortion', with amplitude and shape similar to that described in \citeauthor{GriestK_10a}. The value of $\dmu$ derived from this `corrected' spectrum was $(+3.7\pm5.5)\times10^{-6}$, implying a systematic error on $\dmu$ of approximately $\pm1.9\times10^{-6}$.

Adding these two main systematic error estimates in quadrature, we obtain our final result:
\begin{equation}\label{eq:final_result} 
	\dmu = (+5.6\pm5.5_{\rm stat}\pm2.7_{\rm sys})\times10^{-6}. 
\end{equation} 

\section{Conclusions}

We have measured $\dmu$ from a new H$_2$ absorber $z_{\rm abs}=2.059$, with a large number of well-defined H$_2$ transitions and including some HD transitions, from a very high resolution ($R\approx110000$) Keck/HIRES spectrum. The final value in equation (\ref{eq:final_result}) includes the formal 1-$\sigma$ statistical error and systematic errors estimated from possible long- and short-range wavelength calibration errors. This is the first Keck-based constraint with precision comparable to those obtained from VLT spectra by \citet{KingJ_08a} who use a similar technique to derive $\dmu$. Our new null constraint is consistent with the null constraints from the VLT studies, including those of \citet{ThompsonR_09a} and \citet{WendtM_08a}.

At $z<1$, comparison of the radio inversion transitions of NH$_3$ with molecular rotational lines has yielded two very strong constraints, $\dmu=(0.74\pm0.47_{\rm stat}\pm0.76_{\rm sys})\times10^{-6}$ at $z=0.68$ \citep{MurphyM_08b} and $(+0.08\pm0.47_{\rm sys})\times10^{-6}$ at $z=0.89$ \citep{HenkelC_09a}. While these clearly have superior precision and smaller potential systematic errors than the H$_2$/HD constraints, direct comparison is difficult because of the possibility, in principle, for spatial $\mu$ variations: the different molecular species (NH$_3$ and H$_2$/HD) trace regions of different densities and, therefore, different spatial scales and environments. If $\mu$ does vary, we do not know what that variation depends on, so it is presumptuous to prefer one type of measurement over the other. Clearly, much larger samples of both NH$_3$ and H$_2$/HD constraints, in overlapping redshift ranges, would allow additional tests for systematic errors and stronger conclusions to be drawn.

It is also important to reduce the statistical errors in individual H$_2$/HD measurements by substantially increasing the SNR of the optical spectra. We have demonstrated here that, at least for our spectrum of J2123$-$0050, the statistical error dominates over systematic errors. Nevertheless, if such systematic errors are not also reduced by improving the SNR, it becomes more important to increase the number of H$_2$/HD absorbers for measuring $\dmu$. Surveys targeting particularly gas- and/or metal-rich absorbers have been successful in discovering more H$_2$/HD systems \citep[e.g.][]{NoterdaemeP_08a} but the number known in the northern hemisphere is small.
 
\begin{acknowledgements}
MTM thanks the Australian Research Council for a QEII Research Fellowship (DP0877998). JXP is supported by NSF grants AST-0709235 \& AST-0548180. WU acknowledges financial support from the Netherlands Foundation for Fundamental Research of Matter (FOM).
\end{acknowledgements}


\end{document}